\documentclass[referee]{raa}            

\usepackage{graphicx,times}             
\usepackage[authoryear]{natbib}
\usepackage{amssymb,amsmath}
\bibpunct{(}{)}{;}{a}{}{,}

\usepackage[pagebackref=true]{hyperref}

\usepackage{xcolor} 
\usepackage{ulem}   

\begin{document}

  \title{Fast Radio Burst Cosmology: Hubble Tension and Dark Energy
}

   \volnopage{Vol.0 (20xx) No.0, 000--000}      
   \setcounter{page}{1}          

   \author{Fayin Wang 
      \inst{1,2}
   \and Xuandong Jia
      \inst{1}
   \and Daohong Gao
      \inst{1}
         \and Zigao Dai
      \inst{3}
   }

   \institute{School of Astronomy and Space Science, Nanjing University, Nanjing 210093, China; {\it fayinwang@nju.edu.cn}\\
        \and
             Key Laboratory of Modern Astronomy and Astrophysics (Nanjing University), Ministry of Education, Nanjing 210093, China\\
        \and
             Department of Astronomy, University of Science and Technology of China, Hefei 230026, China\\
\vs\no
   {\small Received 20xx month day; accepted 20xx month day}}

\abstract{Fast radio bursts (FRBs) are luminous, millisecond-duration extragalactic radio transients that have emerged as a powerful, complementary cosmological probe for investigating the late-time cosmic evolution, offering unique advantages over conventional probes such as Type Ia supernovae, baryon acoustic oscillations, and cosmic microwave background radiation. This review systematically summarizes the cosmological applications of FRBs, focusing on their critical roles in measuring the Hubble constant ($H_0$) and constraining dark energy properties. Benefiting from the precise dispersion measure (DM) - redshift relation of localized FRBs, the integrated electron density of the intergalactic medium (IGM) along the line of sight can be tightly modeled, enabling independent and low-redshift measurements of the cosmic expansion rate. Current FRB samples consisting of localized and non-localized events provide competitive $H_0$ constraints, offering an independent method to measure $H_0$. FRBs also serve as effective tracers to constrain dark energy equation-of-state parameters. We comprehensively discuss key limiting factors for FRB cosmological precision, including uncertainties in Galactic and host galaxy electron density models, and IGM inhomogeneities. With the rapid growth of high-precision FRB surveys and localized FRB samples, FRBs are promising to provide stringent constraints on late-time cosmic acceleration, dark energy evolution and cosmic baryons.
}

   \authorrunning{F. Y. Wang, D. H. Gao, X. D. Jia \& Z. G. Dai }            
   \titlerunning{FRB Cosmology: Hubble tension and dark energy}  

   \maketitle
\newpage
\tableofcontents
\newpage
%
%
\section{Introduction}
The standard Lambda cold dark matter ($\Lambda$CDM) model can account for a wide range of astronomical observations. However, serious
challenges to the $\Lambda$CDM model have emerged, among which the most severe ones are the Hubble
constant tension and the deviation from the $\Lambda$CDM model reported by the Dark Energy Spectroscopic Instrument (DESI) collaboration \citep{DESI_DR1,DESI_DR2}.
The Hubble constant tension refers to a statistically significant discrepancy  ($\sim 5\sigma$) between two high-precision measurements of the cosmic current expansion rate $H_0$: the early-universe measurement using the cosmic microwave background (CMB) within the standard $\Lambda$CDM model, which yields $H_0 =67.4 \pm 0.5$ km/s/Mpc \citep{2020A&A...641A...6P}, and the late-universe ``cosmic distance ladder” method (using Cepheid variables and type Ia supernovae, calibrated by the SH0ES team with Hubble and JWST data), which gives $H_0 = 73.04 \pm 1.04$ km/s/Mpc \citep{2022ApJ...934L...7R}. This tension, sharpened by recent independent checks, is possibly due to systematic errors, new physics (e.g., dynamical dark energy, modified gravity), or other fundamental cosmological revisions, making it a major crisis in modern cosmology.

The $\Lambda$CDM model assumes dark energy is a constant cosmological constant, exerting a uniform repulsive force that drives the cosmic accelerating expansion. However, combined analyses of DESI baryon acoustic oscillation (BAO) data with CMB, and type Ia supernovae reveal tentative evidence that the equation of state of dark energy is not $-1$ but evolves over time. Different data combinations yield tension significances ranging from 2.5$\sigma$ to 4.2$\sigma$. This suggests dark energy may be a dynamic field (dubbed ``dynamical dark energy”) rather than a static cosmological constant.

New cosmological probes are urgently required for resolving these challenges. Fast radio bursts (FRBs) are a new type of probe with high precision. FRBs are extremely luminous, transient astronomical phenomena characterized by radio pulses lasting only a few milliseconds. Since the discovery of the first event in archival data from the Parkes telescope \citep{Lorimer2007}, they have rapidly evolved from mysterious anomalies into a major frontier in modern astrophysics. The subsequent detection of a population of highly dispersed bursts \citep{Keane2012, Thornton2013} supported their extragalactic origin, as their observed dispersion measures (DMs) significantly exceeded the maximum contribution expected from the Milky Way's interstellar medium and halo.

A turning moment in FRB research occurred with the discovery of the first repeating source, FRB 121102A \citep{Spitler2016}. The repeating nature of this source attracted extensive follow-up observations using interferometric arrays, which eventually led to the first precise localization of FRB 121102A to a dwarf star-forming galaxy at a redshift of $z = 0.19$ \citep{Chatterjee2017, Tendulkar2017}. This breakthrough not only confirmed the cosmological distances of FRBs but also opened a new era of using them as probes for the distant Universe.

In recent years, the deployment of wide-field radio interferometers, such as the Canadian Hydrogen Intensity Mapping Experiment (CHIME) {\citep{CHIMECollaboration2022}}, the Australian Square Kilometre Array Pathfinder (ASKAP) {\citep{Hotan2021}}, and the Deep Synoptic Array (DSA) {\citep{Law2024}}, has exponentially increased the sample size of FRBs. For instance, the CHIME/FRB project has released comprehensive catalogs containing thousands of bursts from both non-repeating and repeating sources, providing unprecedented datasets with uniform selection effects \citep{CHIME2021, CHIME2023, FRBCollaboration2026}. Simultaneously, continuous monitoring by highly sensitive telescopes like the Five-hundred-meter Aperture Spherical radio Telescope (FAST) has revealed complex bursting behaviors, local environment and energy distributions \citep{Luo2020,Li2021,Xu2022,Wang2022,Li2026}, shedding light on the underlying physical mechanisms.

Although the exact central engines and emission mechanisms of FRBs are still under active debate, with magnetar models currently being the leading scenario \citep{Xiao2021, 2022A&ARv..30....2P, Zhang2023}, their cosmological origin and the rapid accumulation of localized samples make them reliable and independent probes. As observational data continue to grow, FRBs are increasingly utilized across a diverse range of cosmological applications \citep{Bhandari2021, Wu2024}. To be specific, the unique propagation effects encoded in FRB signals have established them as competitive tools for measuring the Hubble constant and the cosmic expansion rate \citep{Wu2020, Hagstotz2022, Wu2022, James2022, Wei2023, Gao2024, Kalita2025,WangYY2025,Xu2025}, as well as for constraining the dark energy equation of state \citep{2014PhRvD..89j7303Z, 2018ApJ...856...65W}. Furthermore, FRBs have played a pivotal role in locating the cosmic missing baryons within the intergalactic and circumgalactic media \citep{Macquart2020, Yang2022, Lin2023, Wang2023, Connor2025}. Beyond these core applications, the growing FRB population is also being exploited to investigate the epoch of helium reionization \citep{Bhattacharya2021, Zhang2021}, place upper bounds on the rest mass of the photon \citep{Wang2021, Lin2023a, Wang2024}, and hunt for compact dark matter objects through gravitational lensing effects \citep{Munoz2016, Wang2018,Laha2020,Zhou2022,GaoR2024}.

This review is organized as follows. In Section 2, we briefly introduce the basic observational properties of FRBs. In Section 3, we discuss the progress on estimating $H_0$ using FRBs. 
We present the constraints on dark energy from FRBs in Section 4. Finally, we give a brief outlook on future prospects.

\section{Basic properties of FRBs}
\label{FRB property}
\subsection{Dispersion Measure}
When electromagnetic waves propagate through a magnetized plasma, their group velocities become frequency-dependent due to interactions with free electrons. This dispersive effect causes lower-frequency radio waves to travel slower than higher-frequency ones, resulting in a characteristic time delay $\delta t$. For an FRB detected at two different frequencies $\nu_1$ and $\nu_2$, the time delay is given by
\begin{equation}
    \delta t = \frac{e^2}{2\pi m_e c} (\frac{1}{\nu_1^{2}}-\frac{1}{\nu_2^{2}})\ \rm{DM}.
\end{equation}
The dispersion measure (DM) is fundamentally defined as the column density of free electrons integrated along the line of sight from the source to the observer:
\begin{equation}
    {\rm DM} = \int \frac{n_e}{1+z} dl,
\end{equation}
where $n_e$ is the free electron number density, $l$ is the physical path length, and the $(1+z)$ factor accounts for the cosmological redshift of the rest-frame frequencies. Because the majority of an FRB's DM originates from outside the Milky Way, it serves as an excellent tracer of the intergalactic electron distribution and the cosmic web.

In practice, the total observed dispersion measure (${\rm DM_{obs}}$) of an FRB is a linear combination of contributions from multiple distinct components along the propagation path:
\begin{equation}
    {\rm DM_{obs}} = {\rm DM_{MW, ISM}} + {\rm DM_{MW, halo}} + {\rm DM_{IGM}} + \frac{{\rm DM_{host}}}{1+z}.
\end{equation}
The Galactic interstellar medium contribution, ${\rm DM_{MW, ISM}}$, is typically estimated and subtracted using established electron density models of the Milky Way, such as NE2001 \citep{Cordes2002}, YMW16 \citep{Yao2017} or the latest model NE2025 \citep{Ocker2026}. The Milky Way halo contribution, ${\rm DM_{MW, halo}}$, is relatively small and can be estimated via cosmological simulations or X-ray observations \citep{Prochaska2019, Keating2020}. 

After removing the Galactic components, the residual extragalactic dispersion measure (${\rm DM_{exc}}$) consists of the IGM and the host galaxy components. Because ${\rm DM_{IGM}}$ encodes the cosmological information while ${\rm DM_{host}}$ depends on the local environment and morphology of the host galaxy, effectively disentangling ${\rm DM_{IGM}}$ from ${\rm DM_{host}}$ remains the primary challenge in utilizing FRBs for high-precision cosmology.

\subsection{The DM of intergalactic medium}
\label{sec:macquart}
The theoretical foundation for utilizing FRBs as cosmological probes rests upon the robust correlation between the dispersion measure contributed by IGM (${\rm DM_{IGM}}$) and the cosmological redshift $z$. Long before a substantial population of FRBs was localized, it was theoretically predicted that the column density of electrons would scale with cosmological distance, offering a novel method to measure cosmic baryons and the expansion history of the Universe \citep{Ioka2003, Inoue2004, Deng2014}.

Assuming a homogeneous and isotropic Universe, the mean expected value of ${\rm DM_{IGM}}$ at a given redshift is expressed through a line-of-sight integral:
\begin{equation}
    \langle {\rm DM_{IGM}}(z) \rangle = \frac{3 c H_0 \Omega_{\rm  b} }{8 \pi G m_{\rm  p}} \int_0^z \frac{f_{\rm IGM}(z')f_{\rm  e}(z')(1+z') }{\sqrt{\Omega_{\rm  m}(1+z')^3 + \Omega_\Lambda}} dz', \label{eq:2.2-1}
\end{equation}
where $c$ is the speed of light, $G$ is the gravitational constant, $m_{\rm  p}$ is the proton mass, and $H_0$ is the Hubble constant. The cosmological parameters $\Omega_{\rm  b}$, $\Omega_{\rm  m}$, and $\Omega_\Lambda$ represent the present-day density fractions of baryons, total matter, and dark energy, respectively. The parameter $f_{\rm IGM}(z)$ denotes the fraction of cosmic baryons residing in the diffuse IGM. The function $f_{\rm  e}(z)$ characterizes the ionization history of the Universe. For FRBs detected at $z < 3$, it is safely assumed that both hydrogen and helium in the IGM are fully ionized, yielding $f_{\rm  e}(z) \approx 7/8$. This theoretical framework was validated by \citet{Macquart2020}, who utilized a sample of localized FRBs from the ASKAP telescope to observationally confirm this tight correlation. Equation~(\ref{eq:2.2-1}) is now widely recognized as the ``Macquart relation''.

\begin{figure}[htbp]
    \centering
    \includegraphics[width=\textwidth]{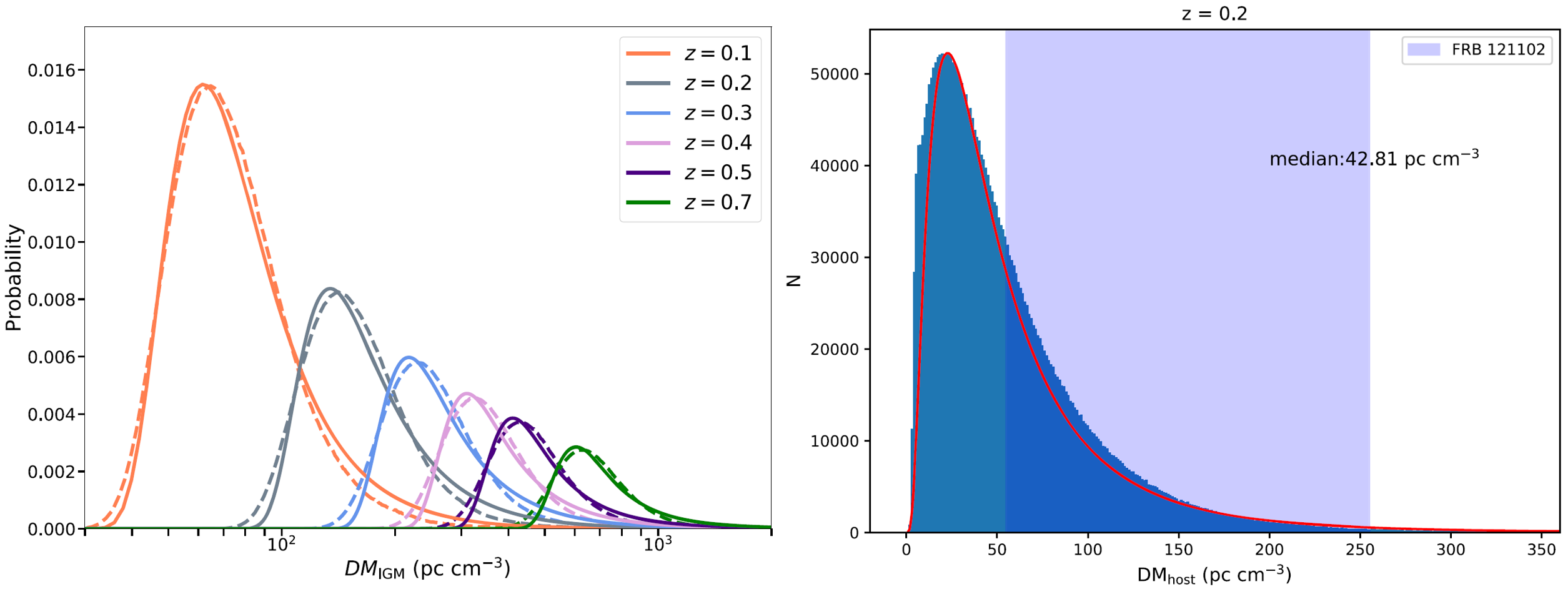}
    \caption{The quasi-Gaussian distribution of $\rm DM_{\rm IGM}$ (left panel) and lognormal distribution of $\rm DM_{\rm host}$ (right panel) from IllustrisTNG simulation. {Dashed lines in the left panel are $\rm DM_{\rm IGM}$ distributions derived from IllustrisTNG simulations and solid lines are the fitting results using Equation~(\ref{eq:2.2-2}). The red line in the right panel shows the best-fitting result of $\rm DM_{\rm host}$ for repeating FRBs like FRB 20121102, and the blue shaded region is the $\rm DM_{\rm host}$ for FRB 20121102 inferred from the observation.} See \citet{Zhang2020} and \citet{Zhang2021} for details.}
    \label{fig:DM_dist}
\end{figure}

However, the actual Universe is not perfectly homogeneous but is instead composed of filaments, clusters, and voids. Consequently, a single FRB line of sight intersects varying large-scale structures, introducing a scatter in the observed ${\rm DM_{IGM}}$ around the mean Macquart relation. The probability density function (PDF) of this scatter is linked to the spatial distribution of intervening baryons and galactic halos \citep{McQuinn2014}. The PDF of ${\rm DM_{IGM}}$ is given by
\begin{equation}
p_{\rm IGM}(\Delta)=A \Delta^{-\beta} \exp \left[-\frac{\left(\Delta^{-\alpha}-C_{0}\right)^{2}}{2 \alpha^{2} \sigma_{\rm IGM}^{2}}\right],\ \Delta=\frac{\rm DM_{IGM}}{\langle{\rm DM_{IGM}}\rangle},\label{eq:2.2-2}
\end{equation}
where $\alpha$ and $\beta$ describe halo density profile, while $A$ and $C_0$ are normalization parameters. To realistically model this inhomogeneous dispersion, extensive cosmological hydrodynamic simulations, such as the IllustrisTNG project, have been widely employed \citep{Zhang2021}. The resulting PDF is typically characterized by a skewed, quasi-Gaussian shape featuring a high-DM tail, which accounts for lines of sight that pass close to dense intervening structures, as shown in the left panel in Fig.~\ref{fig:DM_dist}.

\subsection{The Host Galaxy Contribution to DM}
While the intergalactic medium provides a cosmological signal that scales with distance, the dispersion measure contributed by the FRB host galaxy (${\rm DM_{host}}$) is entirely localized to the source's environment. This component is highly complex and depends on multiple factors, including the morphological type and inclination angle of the host galaxy, as well as the specific location of the progenitor within it. Furthermore, the immediate vicinity of the FRB progenitor can also contribute significantly to the local electron column density \citep{Yang2017,Piro2018, Zhao2021,Wang2025}. 

Because ${\rm DM_{host}}$ cannot be deterministically predicted for individual bursts without deep, multi-wavelength observations, it is typically treated as a statistical distribution. Cosmological hydrodynamic simulations, notably the IllustrisTNG project, have been extensively utilized to track the baryonic mass and free electron distribution within simulated galaxies to model this component \citep{Zhang2020}. These numerical studies reveal that, for a population of FRBs tracking star formation or stellar mass, the probability density function of ${\rm DM_{host}}$ can be well approximated by a log-normal distribution, as shown in the right panel in Fig.~\ref{fig:DM_dist}.

Recent observational and statistical studies have provided deeper insights into the physical drivers of ${\rm DM_{host}}$. For example, \citet{Sharma2024} revealed that FRBs are predominantly hosted by massive galaxies with active star formation, compared to models that assume an unbiased tracking of the cosmic star-formation history. Furthermore, the variance of ${\rm DM_{host}}$ serves as a sensitive tracer for baryonic feedback mechanisms \citep{Leung2025, Sharma2026}. Energetic feedback processes driven by supernovae or active galactic nuclei can significantly redistribute gas from galactic halos into the surrounding intergalactic space \citep{Connor2025}.

In observational practice, radio telescopes measure the total extragalactic dispersion measure, which is the sum of the IGM contribution and the redshift-diluted host contribution (${\rm DM_{exc}} = {\rm DM_{IGM}} + {\rm DM_{host}}/(1+z)$). Because both ${\rm DM_{IGM}}$ and ${\rm DM_{host}}$ possess significant intrinsic scatter and cannot be cleanly separated for any single event, a severe parameter degeneracy emerges \citep{Hagstotz2022, Wu2022, Liu2023}. Specifically, an anomalously high ${\rm DM_{exc}}$ could equally arise from a dense burst environment within the host \citep{Bhardwaj2024}, a line of sight intersecting a massive foreground galaxy cluster \citep{Connor2025}, or a universe with a higher expansion rate ($H_0$) and a larger diffuse baryon fraction ($f_{\rm IGM}$) \citep{Lin2023, Kalita2025}. Breaking this $H_0$--$f_{\rm IGM}$--${\rm DM_{host}}$ degeneracy is the central challenge in FRB cosmology. It strictly requires robust Bayesian inference frameworks that simultaneously evaluate the probability density functions of both the IGM and the host galaxy across a statistically significant sample of localized FRBs \citep{Macquart2020, James2022, Wu2024, Gao2025}.

\subsection{Rotation Measure and $L$-RM Relation}
As linearly polarized radio waves propagate through a magnetized plasma, their plane of polarization experiences a frequency-dependent rotation, a phenomenon known as Faraday rotation. The variation in the polarization angle, $\Delta { \Phi}$, scales with the square of the observational wavelength $\lambda$ as $\Delta { \Phi} = {\rm RM} \lambda^2$. The proportionality factor is the rotation measure (RM), which characterizes the line-of-sight integral of the free electron density $n_e$ multiplied by the parallel component of the magnetic field $B_\parallel$:
\begin{equation}
    {\rm RM}_{\rm obs} = 0.81 \int n_e B_\parallel dl,
\end{equation}
where $n_e$ is in ${\rm cm^{-3}}$, $B_\parallel$ is in $\mu{\rm G}$, and $dl$ is in ${\rm pc}$, yielding the RM in ${\rm rad\,m^{-2}}$. While the dispersion measure essentially counts the column density of free electrons, the RM provides a diagnostic of the magnetic field structure along the propagation path. For repeating FRBs, variations in RM offer critical clues about their dynamic local environments. For instance, extreme and highly variable RM values have been detected in active repeaters like FRB 190520B, strongly suggesting that these sources are embedded in dense, highly magnetized environments, such as a young magnetar wind nebula or a supernova remnant \citep{Piro2018, Zhao2021}.

Building upon these observational characteristics, a profound physical connection between the energetic outputs of FRBs and their local magneto-ionic environments has been revealed, leading to the formulation of the luminosity--rotation measure ($L$-RM) relation \citep{Yang2020, Zhang2025}. This correlation is specifically applicable to a subset of active repeating FRBs that are spatially coincident with compact persistent radio sources (PRSs). Following the initial discovery of the PRS associated with FRB 121102A \citep{Chatterjee2017}, continuous observational efforts have identified similar persistent radio counterparts or strong candidates in several other active repeating systems, such as FRB 190520B \citep{Niu2022} and a growing number of recent discoveries \citep{Ibik2024, Bruni2025, Zhao2026}.

Theoretical models suggest that these PRSs originate from a magnetized, synchrotron-radiating nebula surrounding the central engine of the FRB \citep{Yang2020, Yang2022}. Within this environment, the radio luminosity ($L_\nu$) and the Faraday rotation (quantified by ${\rm RM_{src}}$) are governed by the same medium, and thus tightly coupled by the properties of the nebula \citep{Quataert2000, Zhang2025}. This yields an empirical linear scaling relation, $L_\nu \propto |{\rm RM_{src}}|$\citep{Yang2020, Yang2022}. The astrophysical significance of the $L$-RM relation lies in its potential to transform PRSs into cosmological standard candles. By extracting the intrinsic luminosity of a PRS from the rotation measure of its associated FRB, one can derive the luminosity distance to the source independent of the traditional cosmic distance ladder. Applying this scaling relation offers a model-independent strategy to break the severe degeneracies in cosmological parameter estimation.

\section{Measuring Hubble constant from FRBs}

\subsection{$H_0$ from the Macquart Relation}

\subsubsection{Localized FRBs}

\begin{figure}[t]
    \centering
    \includegraphics[width=0.8\textwidth]{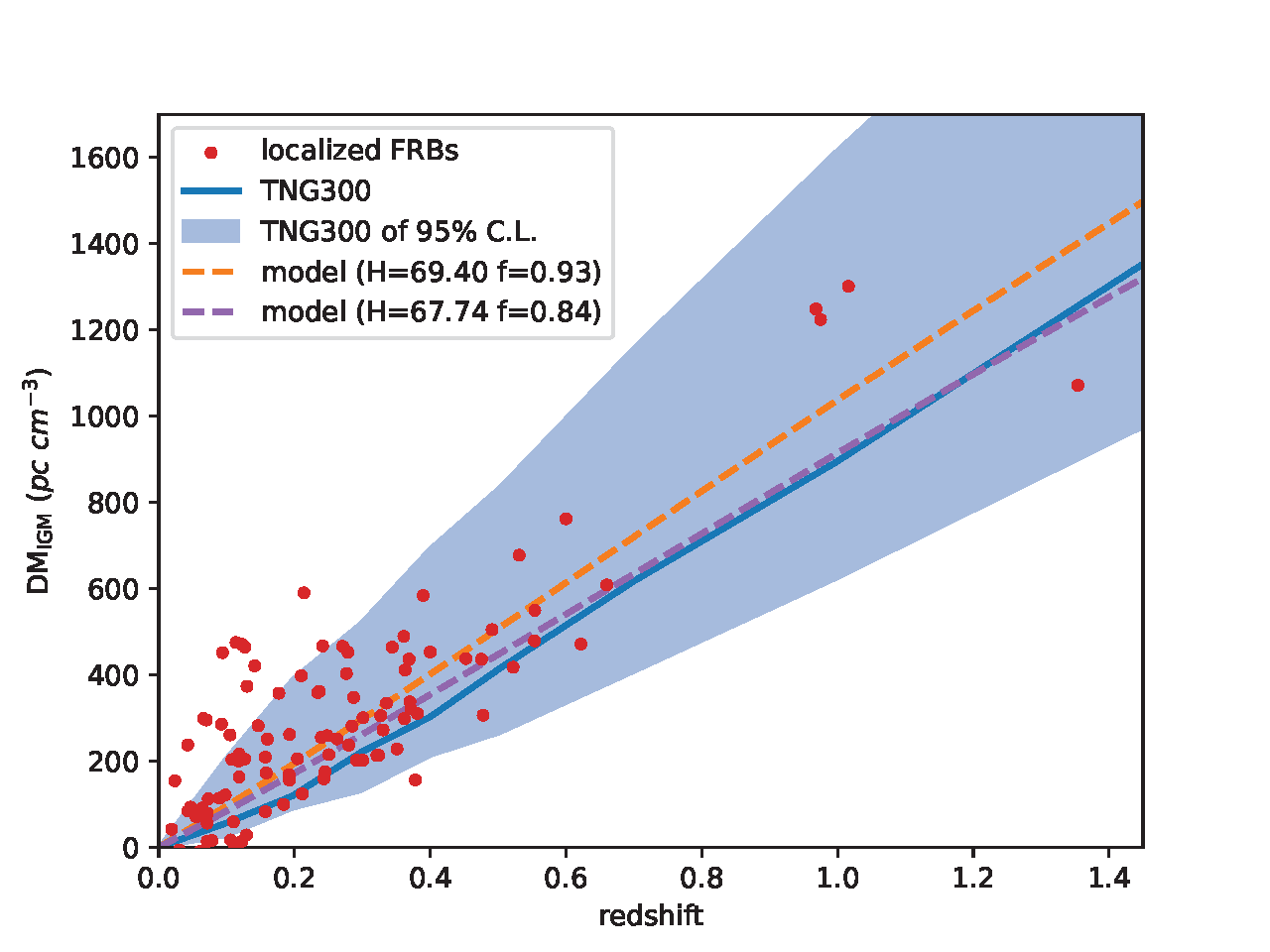}
    \caption{The latest DM-z relation with localized FRBs. The red dots show 108 localized FRBs. The blue line corresponds to the $\rm DM_{\rm IGM}$ from the IllustrisTNG 300 simulation, and the blue shaded area is the 95\% confidence region. The dashed orange and purple lines represent the models in Equation~(\ref{eq:2.2-1}) with different parameters ($H_0$ and $f_{ \rm IGM}$). See \citet{Gao2025} for more details.}
    \label{fig:DM-z}
\end{figure}

The most direct approach to utilizing the Macquart relation for cosmology relies on the unambiguous localization of FRBs to their host galaxies, which provides a precise spectroscopic redshift $z$. Because the mean intergalactic dispersion measure $\langle {\rm DM_{IGM}} \rangle$ is inversely proportional to $H_0$, a sample of localized FRBs can be employed as a novel model-independent probe of the cosmic expansion rate. However, as highlighted in Section~\ref{FRB property}, the observed dispersion measure is a composite quantity. Accurately extracting $H_0$ requires simultaneously handling the highly degenerate and stochastic nature of ${\rm DM_{IGM}}$ and ${\rm DM_{host}}$ \citep{Macquart2020, Zhang2020, Zhang2021,Jahns-Schindler2025}.

Motivated by the growing Hubble tension \citep{Hu2023}, early efforts to measure the Hubble constant were fundamentally limited by the scarcity of localized events, often relying on samples of fewer than 20 FRBs \citep{Hagstotz2022, Wu2022, James2022}. A critical methodological breakthrough during this period was the transition from assuming constant DM contributions to employing probability density functions (PDFs) to disentangle the degenerate IGM and host galaxy components \citep{Macquart2020, Zhang2020, Zhang2021}. By extracting these PDFs from cosmological hydrodynamic simulations (such as IllustrisTNG) or constructing analytical models, researchers developed robust Bayesian and Markov Chain Monte Carlo (MCMC) frameworks to evaluate the dispersion components and constrain $H_0$ \citep{Wu2020, Liu2023, Wei2023}. Although the statistical uncertainties in these pioneering measurements remained relatively large (typically ranging from $8\%$ to $15\%$ \citep{Wu2022, James2022}), they successfully validated the use of FRBs as an independent cosmological probe and laid the analytical groundwork for future analyses \citep{Acharya2025,Fortunato2026,Bao2026}.

\begin{figure}[t]
    \centering
    \includegraphics[width=0.75\textwidth]{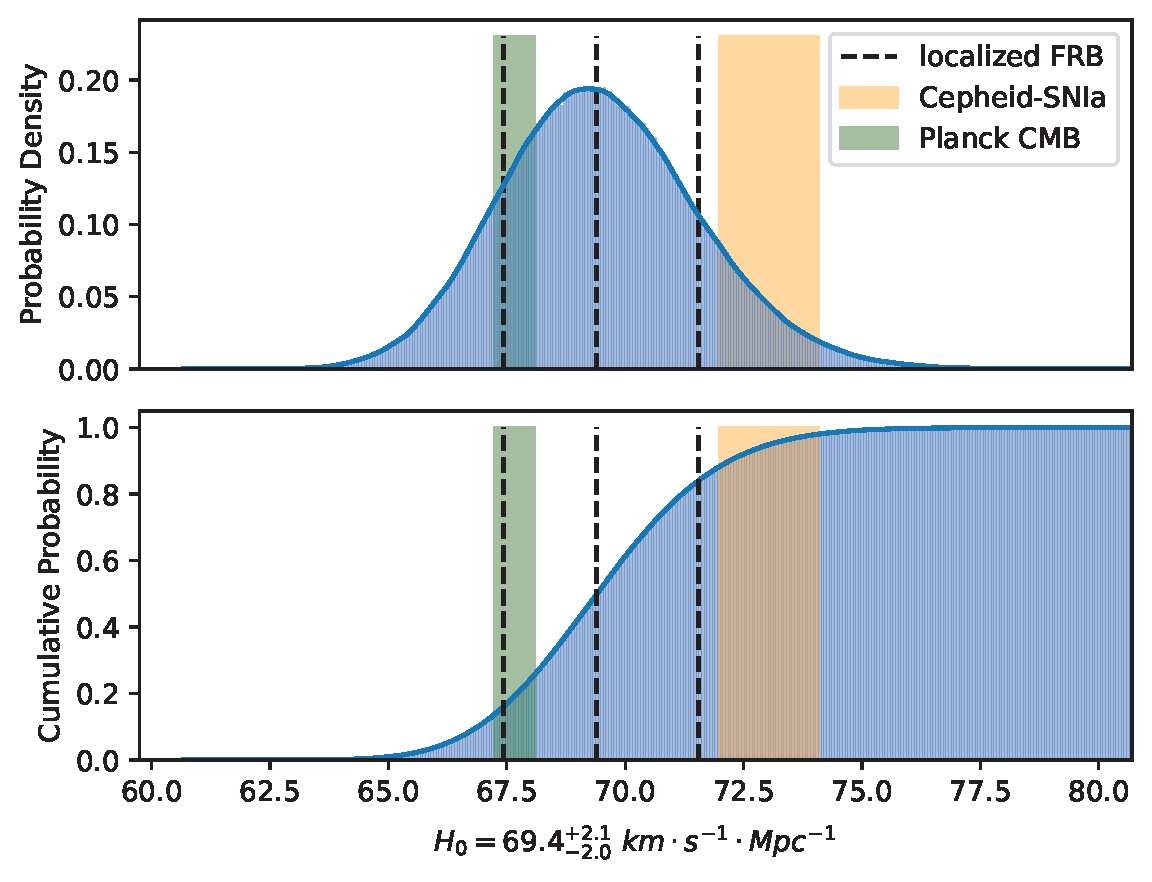}
    \caption{The probability density function and cumulative distribution function of $H_0$ given by 108 localized FRBs. {Dashed lines show the median value and the 68\% confidence interval constrained by localized FRBs. Green and yellow shaded regions are results from early and late cosmological probes.} See \citet{Gao2025} for more details.}
    \label{fig:H0_result}
\end{figure}

Since these foundational studies, the core Bayesian methodology has remained consistent across the literature, while recent advancements are primarily driven by the dramatic increase in the number of localized FRBs and minor specific analytical innovations. Thanks to the enhanced localization capabilities of arrays like ASKAP, DSA, and CHIME/FRB Outriggers, the sample size of host-identified FRBs has rapidly expanded, surpassing one hundred events, as shown in Fig.~\ref{fig:DM-z}. Applying the established MCMC techniques to these expanded datasets has drastically reduced statistical errors \citep{Gao2024,Xu2025, Fortunato2025,2026ApJ...996...50K}. For example, utilizing a sample of 108 localized FRBs, \citet{Gao2025} achieved a refined constraint of $H_0 = 69.4^{+2.1}_{-2.0} \, {\rm km\,s^{-1}\,Mpc^{-1}}$, as shown in Fig.~\ref{fig:H0_result}. \citet{WangYY2025} jointly analyzed 92 localized FRBs alongside Dark Energy Spectroscopic Instrument (DESI) Baryon Acoustic Oscillation (BAO) data to simultaneously constrain the expansion rate and the dark energy equation of state \citep[see also][]{wu2023,Zhang2025c}. Meanwhile, \citet{Zhuge2026} expanded the FRB sample to 117 events, introducing a more accurate analytical probability density function for the diffuse IGM to minimize theoretical biases. Although localized FRBs now provide $H_0$ constraints with statistical uncertainties approaching $\sim 3\%$, the inferred values remain inherently sensitive to assumed priors for the IGM baryon fraction and the strength of galactic feedback. Consequently, overcoming these remaining systematic biases associated with the cosmic baryon distribution will be essential for contemporary analyses to resolve the Hubble tension \citep{Bhandari2021, Wu2024, Hoffmann2024, Sharma2024, Connor2025,Sharma2026b}.

\subsubsection{Non-Localized FRBs}

While localized FRBs provide clean cosmological constraints, the vast majority of detected FRBs, such as the thousands of events cataloged by the CHIME/FRB project \citep{CHIME2021, CHIME2023, FRBCollaboration2026}, lack precise host galaxy identifications. Discarding these unlocalized bursts means wasting a tremendous amount of statistical power. Consequently, various innovative methods have been developed to utilize the unlocalized FRB population.

One intuitive approach is the pseudo-redshift method, which infers a redshift distribution for each unlocalized FRB based on its observed extragalactic dispersion measure \citep{Hackstein2020,Zhang2021b,Qiang2022,Tang2023}. {The physical foundation of this derivation relies on the Macquart relation (as detailed in Section~\ref{sec:macquart}). However, because ${\rm DM_{exc}}$ includes highly uncertain contributions from both the IGM and the local host galaxy, a direct one-to-one mapping from DM to redshift is impossible. Instead, this relation can be inverted in a statistical approach.} By adopting probability density distributions for the IGM and host galaxy DM components, the Macquart relation can be inverted via Monte Carlo sampling to generate a pseudo-redshift posterior for each burst. For instance, \citet{Gao2025} applied this technique to 527 unlocalized CHIME FRBs, demonstrating that expanding the sample size with unlocalized bursts can compress the statistical uncertainty on $H_0$ to the $\sim 1\%$ level. Another compelling strategy borrowed from gravitational wave cosmology is the ``dark siren'' method. The sightlines of FRBs can be spatially cross-correlated with comprehensive galaxy catalogs, such as the DESI Legacy Imaging Surveys. By assigning a host probability to all galaxies within the localization volume (often weighted by galaxy properties like stellar mass or luminosity) and marginalizing over these candidates, $H_0$ can be constrained without requiring a unique host identification \citep{Zhao2022}.

Alternatively, rather than assigning individual host galaxies or pseudo-redshifts, some studies model the entire observed DM distribution of the unlocalized FRB population. By calculating the intrinsic FRB luminosity function, volumetric rate, and selection biases, researchers can marginalize over all plausible redshifts to extract cosmological parameters \citep{Liu2026}. While all these approaches achieve higher statistical precision, their ultimate accuracy remains sensitive to systematic assumptions, underlining the necessity of combining the statistics of unlocalized bursts with the clean anchors provided by localized samples.
\subsection{$H_0$ from Strong Lensing FRBs}
In addition to the dispersion measure-based techniques, the phenomenon of strong gravitational lensing offers a purely geometric and independent pathway to measure the Hubble constant using FRBs \citep{Li2018}. When the propagation path of an FRB is intercepted by a massive foreground object such as a galaxy or a galaxy cluster, the immense gravitational field deflects the radio waves, potentially generating multiple images of the same burst. Because the radiation forming each image travels along a different path and traverses different gravitational potentials, the respective pulses arrive at the observer at different times. 

The arrival time difference between any two lensed images, $\Delta t$, is governed by the time-delay distance $D_{\Delta t}$ and the Fermat potential difference $\Delta { \Phi}$ evaluated between the image positions:
\begin{equation}
    \Delta t=\frac{\Delta \Phi}{c} D_{\Delta t}=\frac{\Delta \Phi}{c} \cdot\left(1+z_{ \rm l}\right) \frac{D_{ \rm l} D_{ \rm s}}{D_{ \rm ls}},
\end{equation}
where $z_{ \rm l}$ is the redshift of the lens, and $D_{ \rm l}$, $D_{ \rm s}$, and $D_{ \rm ls}$ denote the angular diameter distances from the observer to the lens, the observer to the source, and the lens to the source, respectively. Because all angular diameter distances are inversely proportional to the present-day expansion rate, a precise measurement of $\Delta t$, combined with a robust mass model of the lens galaxy to determine $\Delta { \Phi}$, directly yields $H_0$ \citep{Gao2022}.

This technique, widely known as time-delay cosmography, has traditionally been applied to strongly lensed quasars and supernovae. However, extracting time delays from these sources requires cross-correlating broad light curves over months or years, which inherently introduces temporal uncertainties on the order of days. FRBs, in contrast, feature transient pulses with millisecond intrinsic widths. The time delay between lensed FRB images can be extracted with unparalleled precision, frequently down to fractions of a millisecond \citep{Wucknitz2021}. Theoretical forecasts have demonstrated that with ten precisely modeled strongly lensed FRB systems, $H_0$ could be constrained to a sub-percent level, making it a powerful tool to independently resolve the Hubble tension without relying on the IGM baryon fraction \citep{Li2018}. 

While definitive observations of macroscopically lensed FRB systems remain observationally challenging to confirm, promising candidates with repeated, identically shaped bursts with fixed temporal separations have recently been identified in large survey databases{. For instance, FRB 20190308C, FRB 20190320B and FRB 20190627B are identified as lensed candidates with different methods \citep{Krochek2022, Chang2025, Xiong2026}. Furthermore, theoretical forecasts have consistently demonstrated that the exceedingly high all-sky event rate of FRBs will inevitably yield a substantial strongly lensed population \citep{Munoz2016, Wang2018, Gao2022}. As next-generation facilities with unprecedented wide-field sensitivity come online, such as the Deep Synoptic Array (DSA-2000), the Canadian Hydrogen Observatory and Radio-transient Detector (CHORD), and the Square Kilometre Array (SKA), the detection of hundreds of strongly lensed FRB systems is robustly anticipated, which will fully unlock the potential of FRB time-delay cosmography for measuring the Hubble constant \citep{Connor2023, Chang2025}.

\begin{figure}[htbp]
    \centering
    \includegraphics[width=0.9\textwidth]{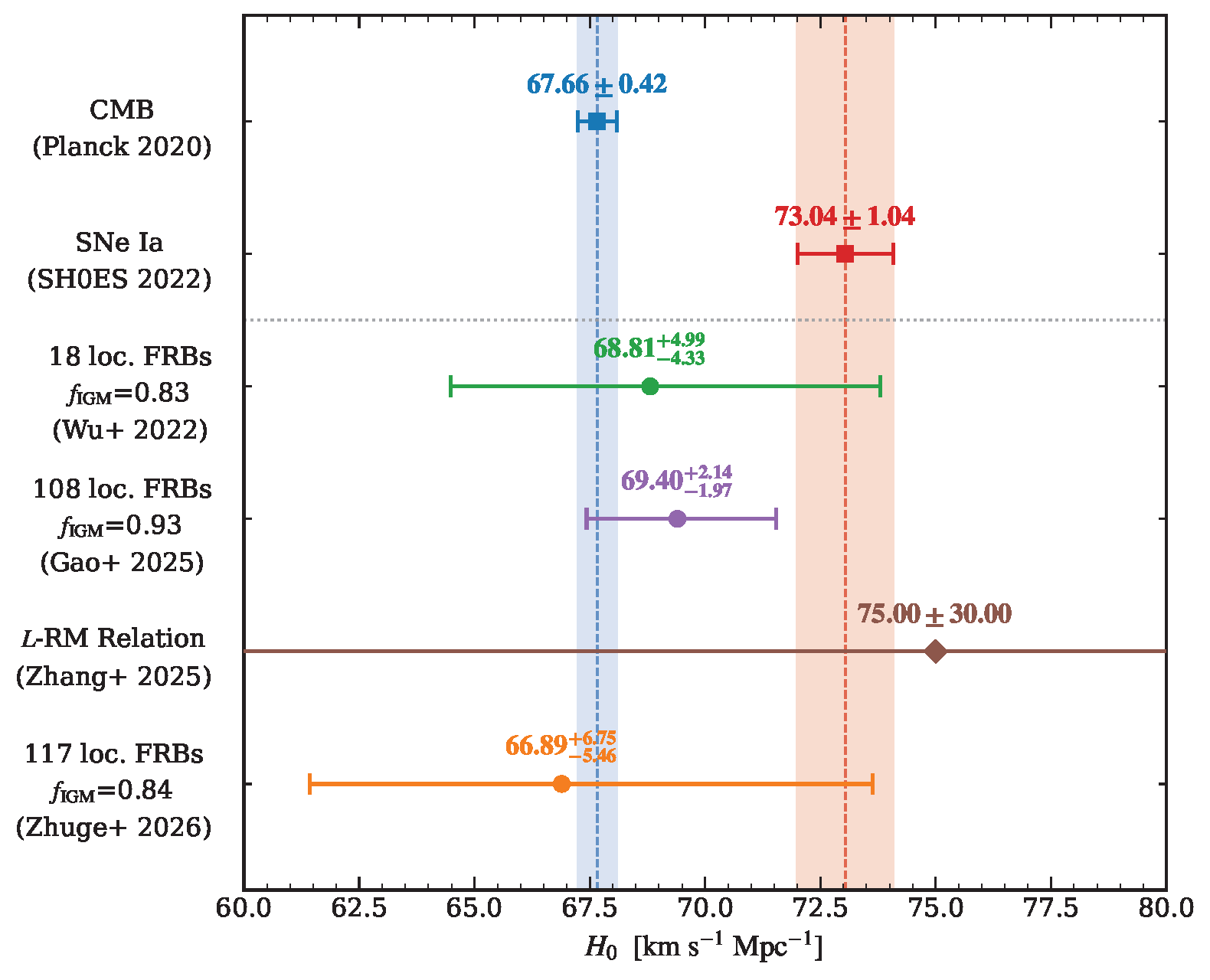}
    \caption{Summary of representative works using FRBs to constrain $H_0$ with different methods and datasets \citep{Wu2022,Gao2025,Zhang2025,Zhuge2026}, as well as early-time measurements using CMB \citep{2020A&A...641A...6P} and late-time measurements using SNe Ia \citep{2022ApJ...934L...7R}.}
    \label{fig:summary}
\end{figure}

\subsection{$H_0$ from the $L$-RM Relation}
The empirical $L$-RM relation connects the intrinsic radio luminosity $L_\nu$ of a PRS to the rotation measure of its associated FRB \citep{Yang2020}, effectively transforming PRS-associated FRBs into standard cosmological candles. By concurrently measuring the observed flux density $F_\nu$ and the local RM of a PRS system, researchers can deduce its intrinsic luminosity and directly compute the luminosity distance $D_L$ to the host galaxy. 

Because the luminosity distance is purely a geometrical function of the source redshift and the cosmic expansion history (governed by $H_0$, $\Omega_{\rm  m}$, and $\Omega_\Lambda$), this method offers a novel pathway to constrain the Hubble constant. Crucially, this approach is fundamentally independent of the Macquart relation. Therefore, it completely avoids the parameter degeneracies caused by the IGM baryon fraction $f_{\rm IGM}$ and the stochastic ${\rm DM_{host}}$ that severely limit the precision of traditional DM-based distance measurements \citep{Kumar2019, Zhang2025}.

Several groups have recently applied the $L$-RM standard candle framework to constrain $H_0$. For instance, \citet{Gao2025b} utilized a sample of four well-calibrated FRB-PRS systems to derive an independent Hubble constant measurement of $H_0 = 86.18^{+18.03}_{-14.99} \, {\rm km\,s^{-1}\,Mpc^{-1}}$. Concurrently, \citet{Zhang2025} employed six confirmed PRS systems, obtaining a preliminary constraint of $H_0 = 75 \pm 30 \, {\rm km\,s^{-1}\,Mpc^{-1}}$. While the current statistical uncertainties remain broad due to the small sample size of confirmed PRS events and their dynamic magneto-ionic environments \citep{AnnaThomas2023}, these pioneering results demonstrate that the method is remarkably robust against systematic uncertainties originating from IGM baryon fraction assumptions. In Fig.~\ref{fig:summary}, we summarize some representative works that employ FRBs in solving Hubble Tension.

\section{Constraints on the evolution history}
To address the Hubble tension, various attempts have been proposed. They can be broadly categorized into three classes: modifications to the early Universe, the late Universe, and the local Universe \citep{2021CQGra..38o3001D,2025PDU....4901965D}. Among these, late Universe modifications seek to explain the larger $H_0$ value at low redshifts by altering the cosmic expansion history, for instance, through modifications to the dark energy equation of state (EoS) \citep{Hu2023,2024JCAP...10..035G,2024PDU....4501539N}. Recently, results from the DESI collaboration based on BAO measurements have indicated a deviation from the standard cosmological model \citep{DESI_DR1,DESI_DR2}. When combined with different SNe datasets, the significance reaches up to $4.2 \sigma$. The deviation has attracted considerable attention and prompted extensive research \citep{2024PhRvD.110l3533P,2025MNRAS.542.1063H,Jia2025_2,2025ApJ...979L..34J,2025PhRvD.112h3528L,2025MNRAS.542L..24C,2025PhRvD.111l3504P,2025IJMPD..3450061P,2025PhRvL.134r1002Y,2026JHEAp..4900428O}. In this section, we present the research on dark energy using FRBs.

\subsection{Simulating FRBs}

The DM of the FRBs encodes the integrated information of the cosmic distance it traverses, which provides an opportunity to use FRBs to constrain cosmological parameters \citep{2014PhRvD..89j7303Z}. Within the framework of the $\Lambda$CDM cosmological model, a cosmological constant dark energy is assumed, with its energy density $c^2 \rho_{\mathrm{DE}}$ remaining constant in both space and time. If the equation of state for dark energy varies with redshift as $w(z) = P(z)/(c^2 \rho_{\mathrm{DE}})$, where $P(z)$ is its pressure, then the energy density evolves as 
\begin{equation}
    \frac{\rho_{\mathrm{DE}}(z)}{\rho_{\mathrm{DE},0}} = \exp \left[ 3 \int_{0}^{z} [1+w(z')] \frac{dz'}{1+z'} \right].\label{eq:4.1-1}
\end{equation}
For the $\Lambda$CDM model, $w$ corresponds to $-1$. For the $w$CDM model, equation (\ref{eq:4.1-1}) is typically expressed as $(1+z)^{3(1+w)}$. Another commonly used parameterization is the CPL model \citep{2003PhRvL..90i1301L}, which is typically expressed as
\begin{equation}
    w(z) = w_0 +w_a \frac{z}{1+z}. \label{eq:4.1-2}
\end{equation}
The model captures the behavior of many physically motivated dark energy models through a simple parameterization. In this case, the mean expected value of DM$_{\mathrm{IGM}}$ from equation (\ref{eq:2.2-1}) becomes
\begin{equation}
    \langle {\rm DM_{IGM}}(z) \rangle = \frac{3 c H_0 \Omega_{\rm  b} }{8 \pi G m_{\rm  p}} \int_0^z \frac{f_{\rm IGM}(z')f_{\rm  e}(z')(1+z') }{\sqrt{\Omega_{\rm  m}(1+z')^3 + \Omega_{\rm DE} (1+z')^{3[1+w(z')]}}} dz', \label{eq:4.1-3}
\end{equation}

Currently, the number of localized FRBs is only around one hundred, which is still far fewer than the thousands of SNe samples accumulated over the long term. Therefore, many researchers currently have to rely on simulated FRB data to evaluate the potential as dark energy probes \citep{2014ApJ...788..189G,2018ApJ...856...65W,2020ApJ...903...83Z,2022JCAP...02..006Q,2023JCAP...04..022Z}. The redshift distribution of simulated data is typically generated based on the FRB event rate. Recent studies using the Lynden-Bell's $C^-$ method have shown that the comoving FRB formation rate decreases with increasing redshift, following $\rho(z) \propto (1+z)^{-5.38 \pm 0.02}$ \citep{2024ApJ...973L..54C,2026ApJ..1003..179J}. \cite{2018ApJ...856...65W} used $1,000$ simulated FRB samples, and the results showed that FRBs provide weak constraints on dark energy. \cite{2020ApJ...903...83Z} predicted that at least $10,000$ well-localized FRBs are required to achieve a measurement precision comparable to CMB and BAO. 

One challenge for FRB cosmology is the measurement precision of $\langle {\rm DM_{IGM}}(z) \rangle$. To serve as a probe complementary to, or even superior to, SNe, the $\langle {\rm DM_{IGM}}(z) \rangle$ measurement precision should be within $10 \%$. While the measurements of individual FRBs fluctuate significantly, observing a sufficient number of FRBs within a narrow redshift range allows the average to cancel out the line-of-sight inhomogeneities. This improves the constraining power of FRBs. Through Monte Carlo simulations, \cite{2014PhRvD..89j7303Z} calculated that if there are $\sim 20$ FRBs per redshift bin, the precision can be constrained to within $10 \%$; with $\sim 80$ FRBs per redshift bin, the precision can be constrained to within $5 \%$.

Despite the currently limited constraints from FRBs, they can be jointly analyzed with other probes \citep{2020ApJ...903...83Z,2026PhRvD.113b3537F}. Both theoretical models and observations suggest that a small fraction of FRBs and GRBs might be associated. If such FRB/GRB association systems are observed in the future, their redshifts can be determined via GRB afterglow observations, while the DM can be derived from the FRBs. \cite{2014ApJ...788..189G} predicted that even just a few dozen associated FRB/GRB events could constrain the EoS parameter $w$ as effectively as hundreds of SNe. In the upcoming era of the SKA, large samples of FRB data are expected to deliver more precise cosmological measurements \citep{2023SCPMA..6620412Z}, and hold great promise for probing dark energy \citep{2023JCAP...04..022Z,2025ChPhC..49k5109Y}.

\subsection{Possible redshift dependent $H_0$}
To date, numerous theories have been proposed to resolve or alleviate the Hubble tension, including \citep{2021CQGra..38o3001D,2021A&ARv..29....9S,2022NewAR..9501659P,2025PDU....4901965D}. Recently, studies on gravitational lensing of quasars have indicated that the Hubble constant inferred from strongly-lensed quasar time delay (H0LiCOW) exhibits a mild decline with lens redshift \citep{2020MNRAS.498.1420W}. The trend is statistically modest, with a significance of approximately $1.9 \sigma$. The inclusion of an additional new $H_0$ measurement \citep{2020MNRAS.494.6072S} further reduces the significance to $1.7 \sigma$ \citep{2020A&A...639A.101M}, as illustrated in Fig. \ref{fig:4.2-1}. It is worth noting that although this declining trend is statistically notable, time-delay cosmology still presents a gap compared to the most widely used SNe cosmology. Benefiting from decades of observations, current SNe samples have accumulated to several thousand, which helps reduce statistical errors and improve the precision of cosmological constraints \citep{SNe_pantheon+,SNe_DESY5,SNe_Union3}. However, unlike SNe samples, which rely on the calibration of absolute magnitudes, quasar gravitational lensing is better able to provide independent measurements at different redshifts. Although we cannot rule out the possibility that the declining trend is caused by some unknown systematics, it inspires a new perspective for explaining the Hubble tension.

\begin{figure}[htbp]
    \centering
    \includegraphics[width=0.8\textwidth]{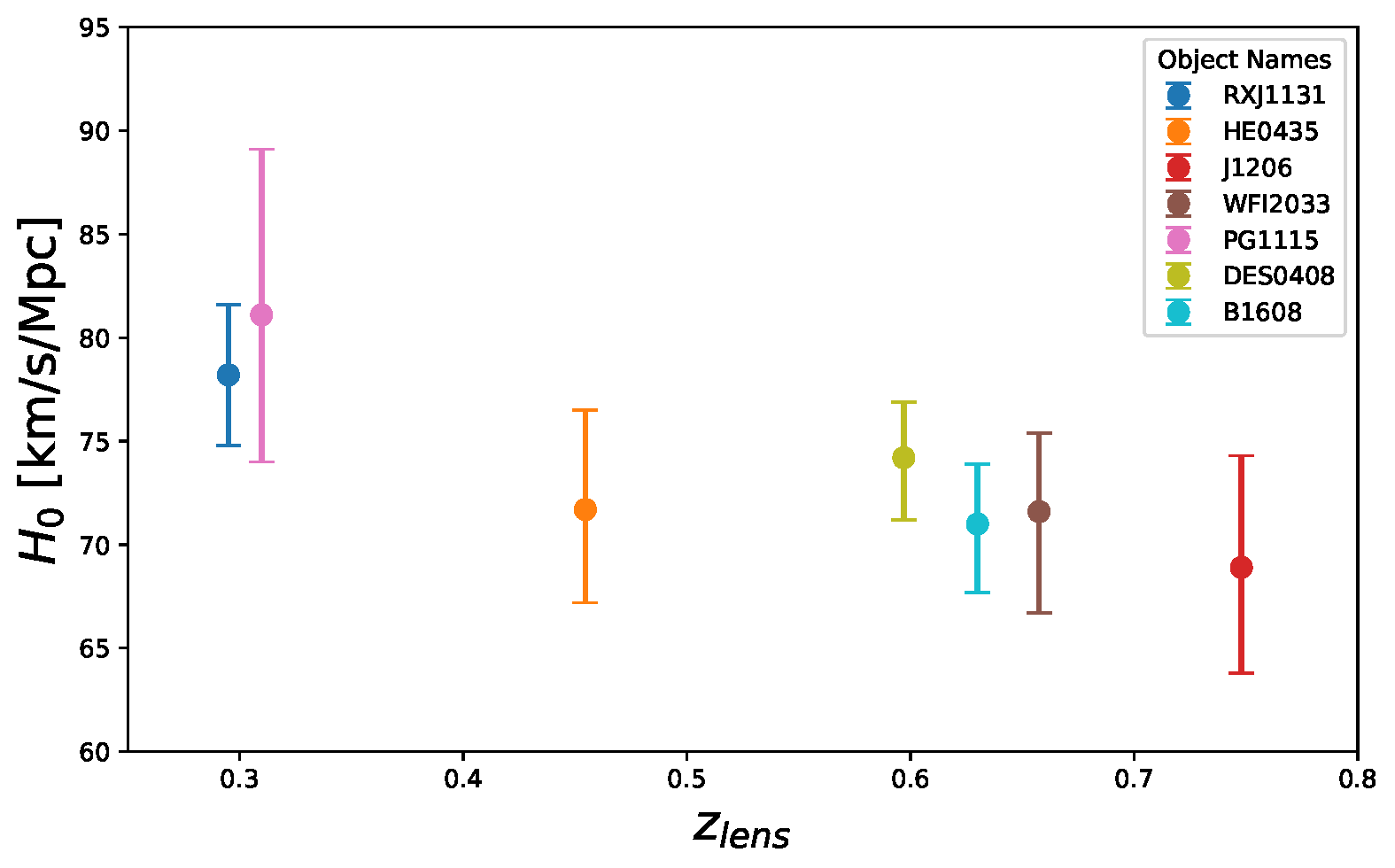}
    \caption{$H_0$ measurements from H0LiCOW. The declining trend of $H_0$ value with increasing lens redshift has significance levels of $1.7 \sigma$. More details can be seen in Ref. \cite{2020A&A...639A.101M}.}
    \label{fig:4.2-1}
\end{figure}



The possible evolution of the Hubble constant has attracted significant attention, prompting numerous fascinating investigations. Through direct binning analysis of the data, a similar declining trend was identified with a confidence level of approximately $2.1 \sigma$ \citep{2020PhRvD.102j3525K}. Furthermore, within the framework of $\Lambda$CDM and $w$CDM models and assuming a prior for the evolution of $H_0$, a similar declining trend was obtained from the data analysis \citep{2021ApJ...912..150D}. They investigated different binning methods as well as cosmological models, and concluded that the declining trend of $H_0$ is robust, being independent of both the model and the binning scheme. Within the framework of the $\Lambda$CDM model, they obtained a local $H_0 = 73.493 \pm 0.144$ km s$^{-1}$ Mpc$^{-1}$, and extrapolated it to $H_0(z = 1100) = 69.271 \pm 2.815$ km s$^{-1}$ Mpc$^{-1}$ at the CMB epoch under the assumption of evolution. It alleviates the Hubble tension by approximately $66 \%$. Although these two studies were conducted independently, their results corroborate each other. This provides stronger support for the hypothesis that $H_0$ exhibits redshift-dependent evolution, and has attracted increasing attention. Follow-up research yields similar results, corroborating the declining trend observed in the Hubble constant \citep{2021PhRvD.103j3509K,2022A&A...668A..34H,2022PhRvD.106d1301O,2022Galax..10...24D,2024PDU....4401464O,Dai:2026pvx}.

Following the preliminary investigations into the potential evolution of $H_0$, some studies have delved deeper to examine the specific form or functional behavior of its evolution \citep{2023A&A...674A..45J,2025ApJ...979L..34J}. Researchers aim to infer $H_0$ directly from the data, without relying on model assumptions. However, if the dataset is simply partitioned and fitted within each bin, the resulting estimates of $H_0$ often exhibit a certain degree of degeneracy. Since distance is inherently a cumulative function, it is evident that fitting $H_0$ in high-redshift bins inherently depends on the values of $H_0$ at low redshifts. Conversely, assuming a specific model or a prescribed form of $H_0$ evolution inevitably introduces additional prior dependence into the results, thereby biasing the final inference. Therefore, obtaining unbiased and mutually independent estimates of $H_0$ at different redshifts remains a challenging issue.

To address the aforementioned issues, a non-parametric approach was pioneered to infer the values of $H_0$ at different redshifts \citep{2023A&A...674A..45J}. Inspired by studies on the dark energy equation of state \citep{2005PhRvD..71b3506H}, they proposed a similar approach to investigate the evolution of $H_0$ in a non-parametric way. The method is based on the assumption that the Hubble constant within a given redshift bin can be approximated as a constant. Therefore, the variation of the Hubble constant can be represented by a step function:
\begin{equation}\label{eq:H0function}
H_0(z)=\begin{cases} H_{0,z_1} &\text{ if } 0\le z < z_1, \\ 
H_{0,z_2} &\text{ if } z_1 \le z < z_2,\\
\cdots  &\cdots,\\
H_{0,z_i} &\text{ if } z_{i-1} \le z < z_i,\\
\cdots &\cdots, \\
H_{0,z_N} &\text{ if } z_{N-1} \le z < z_N.
\end{cases}
\end{equation}
The parameter $i$ means the $i$th redshift bin, $N$ is the number of total redshift bins, and $H_{0,z_i}$ represents the value of $H_0(z)$ between $z_{i-1}$ and $z_i$. If $H_0(z)$ is evolving, the formula can approximately yield the values of $H_0(z)$ across different redshift bins. As $N$ increases, the number of bins grows, and the results will approach the true values. Conversely, if $H_0(z)$ is constant, the estimates of $H_0$ across different redshift bins will remain consistent within the error margins, effectively recovering the true $H_0$. 
By combining this parametrization with a cosmological model, one can obtain an extended evolutionary form of $H_0(z)$ within that framework. The simplest case arises in the standard cosmological model, where $H(z)$ can be expressed as:
\begin{equation}\label{eq:Hz step}
\begin{split}
H(z_i)=& H_{0,z_1}\int_{0}^{z_1} \frac{3\Omega_{m0}(1+z)^2}{2\sqrt{\Omega_{m0}(1+z)^{3}+\Omega_{\Lambda0}}} \\
& +H_{0,z_2}\int_{z_1}^{z_2} \frac{3\Omega_{m0}(1+z)^2}{2\sqrt{\Omega_{m0}(1+z)^{3}+\Omega_{\Lambda0}}} \\
& +\cdots \\
& +H_{0,z_i}\int_{z_{i-1}}^{z_i} \frac{3\Omega_{m0}(1+z)^2}{2\sqrt{\Omega_{m0}(1+z)^{3}+\Omega_{\Lambda0}}}+H_{0,z_i}.
\end{split}
\end{equation}
If $H_0(z)$ does not exhibit any evolutionary trend, the result will revert to $H_0$, as shown in Fig. \ref{fig:4.2-5}. It is evident that when $H_0$ is constant, the $H(z)$ derived from Equation (\ref{eq:Hz step}) exactly matches the prediction of the $\Lambda$CDM model. 

\begin{figure}[htbp]
\centering
\includegraphics[width=0.8\textwidth]{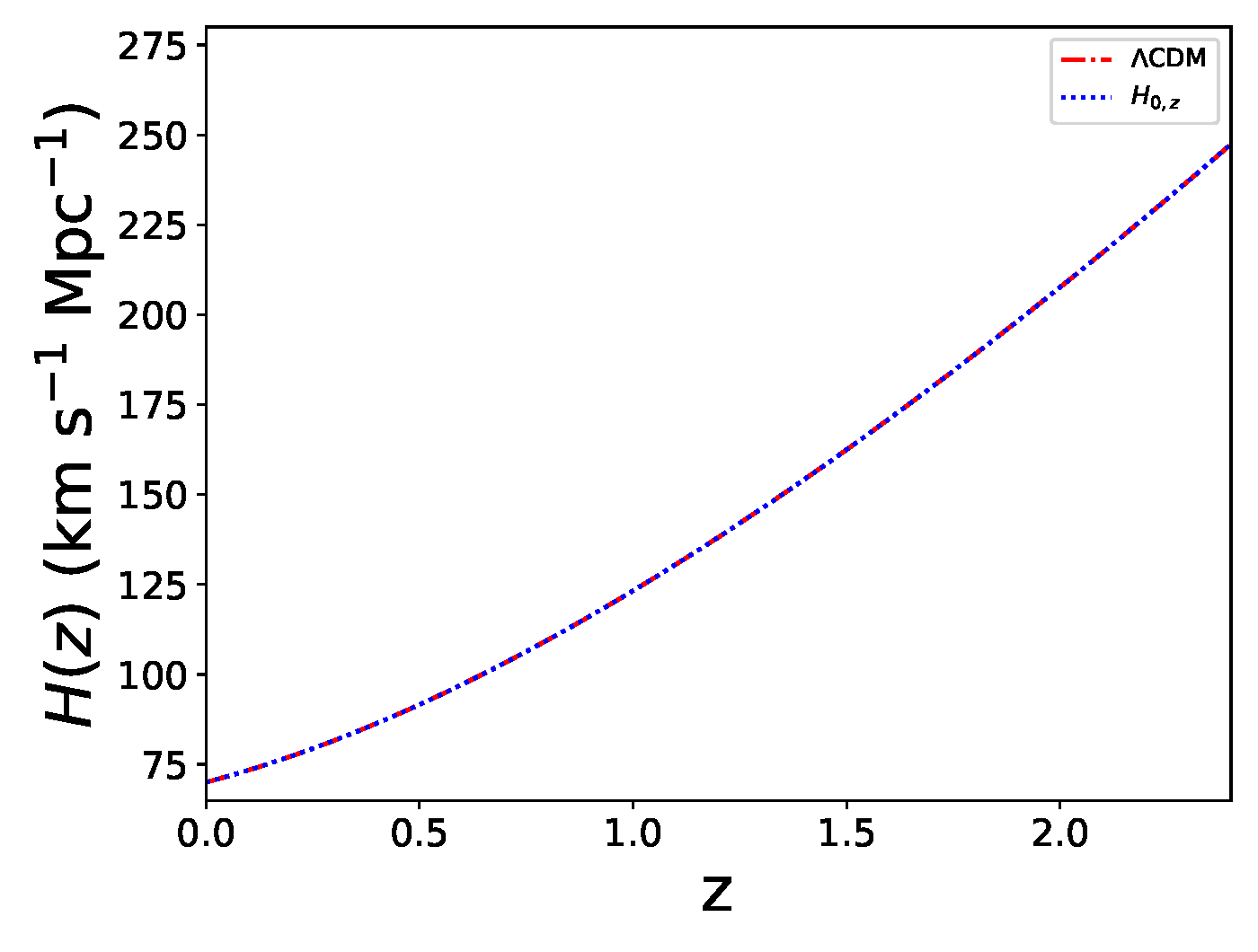}
\caption{Comparison between the Hubble parameter $H(z)$ in the standard $\Lambda$CDM model and that derived from Equation (\ref{eq:Hz step}). $H_0=H_{0,z_i} = 70$ km s$^{-1}$ Mpc$^{-1}$, $\Omega_{k0}=0,$ and $\Omega_{m0} = 0.3$ are assumed. More details can be seen in Ref. \cite{2023A&A...674A..45J}. 
\label{fig:4.2-5}}
\end{figure}

The posterior distribution of $H_0$ was derived by performing an MCMC fit to the combined SNe, BAO, and CC data. However, the inferred $H_0$ values across different redshift bins are not statistically independent. To remove the degeneracy, a transformation matrix was calculated following \cite{2005PhRvD..71b3506H}. Principal Component Analysis (PCA) was then employed to compress the observational constraints while retaining the full information content. The evolution of $H_0$ was examined across diverse samples and binning schemes. The analysis reveals that the observed decline trend in $H_0$ is robust and independent of the specific binning strategy or sample selection. Specifically, using equal-width binning, a significant $5.6 \sigma$ declining trend was detected, as illustrated in Fig. \ref{fig:4.2-6}. This data-driven result reveals a clear evolution of $H_0$ with redshift: at low redshifts, the inferred $H_0$ aligns with the local distance ladder value \citep{2022ApJ...934L...7R}, while at high redshifts, it converges toward the value inferred from CMB measurements \citep{2020A&A...641A...6P}. The redshift-dependent behavior effectively alleviates the current Hubble tension, providing a possible explanation.

With the availability of additional observational data, a more detailed and refined analysis was conducted \citep{2025ApJ...979L..34J}. By incorporating the latest BAO measurements and updated SNe Ia samples, the analysis yielded a consistent declining trend in $H_0$, in agreement with earlier findings \citep{2025ApJ...979L..34J}. The declining trend has been consistently confirmed across different binning schemes and independent observational samples, reinforcing its robustness.

\begin{figure}[htbp]
\centering
\includegraphics[width=\textwidth,angle=0]{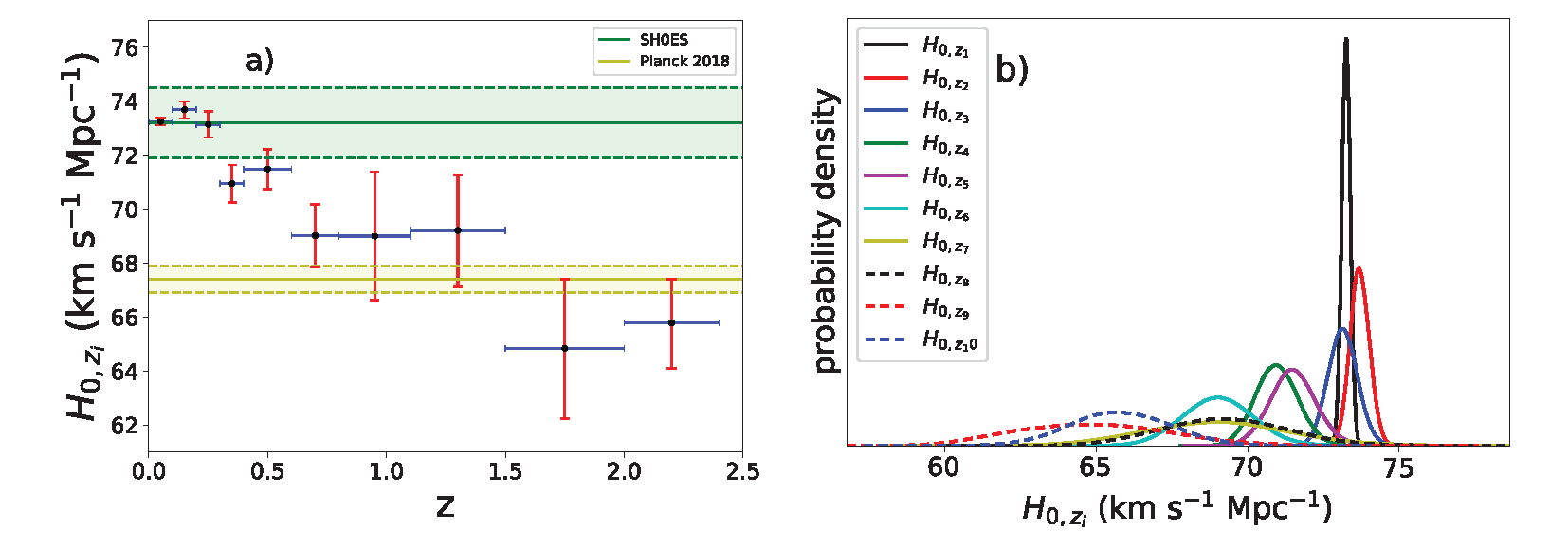}
\caption{Fitting results for $H_0(z)$ using equal-width binning across ten redshift intervals. Left panel: $H_0(z)$ as a function of redshift. A clear decreasing trend is observed, with a significance of $5.6\sigma$ at $z>0.3$. Right panel: Normalized probability distributions of $H_0(z)$ for ten redshift bins, which are well-approximated by Gaussian profiles. More details can be seen in Ref. \cite{2023A&A...674A..45J}.}
\label{fig:4.2-6}
\end{figure}

Another popular model-independent approach is Gaussian Process (GP). It is widely used in cosmological research \citep{2018ApJ...856....3Y,2018JCAP...04..051G,2019ApJ...886L..23L,2020ApJ...895L..29L,2022PhyS...97h5011S,2021PhRvD.103j3509K,2021MNRAS.507..730H}.
GP allows one to obtain a continuous and smooth function from discrete measurements of the cosmic expansion history. The reconstructed function can be used to estimate the value of $H(z)$ at any redshift within the range, including the current Hubble constant $H_0$, which corresponds to the value at $z = 0$. 
Unlike previous studies, \cite{2022MNRAS.517..576H} proposed a new grouping method to investigate the variation of the Hubble constant. They adopted a cumulative approach, progressively incorporating CC and BAO data into the research sample. In this way, the variation in $H_0$ obtained from each GP reconstruction reflects the impact of the newly added data. Their results reveal a transition in $H_0$ during the late Universe, as shown in Fig. \ref{fig:4.2-4}. As redshift increases, $H_0$ gradually decreases from a high value at low redshift to a low value at high redshift. This trend of redshift-dependent $H_0$ evolution helps to alleviate the Hubble tension, with the transition point occurring around $z \approx 0.49$. Without introducing any new cosmological model, this result offers a potential solution to the Hubble tension, achieving a mitigation level of approximately $70 \%$, and remains consistent with $H_0$ measurements from H0LiCOW within the $1 \sigma$ confidence level. 

\begin{figure}
\includegraphics[width=6.8 cm]{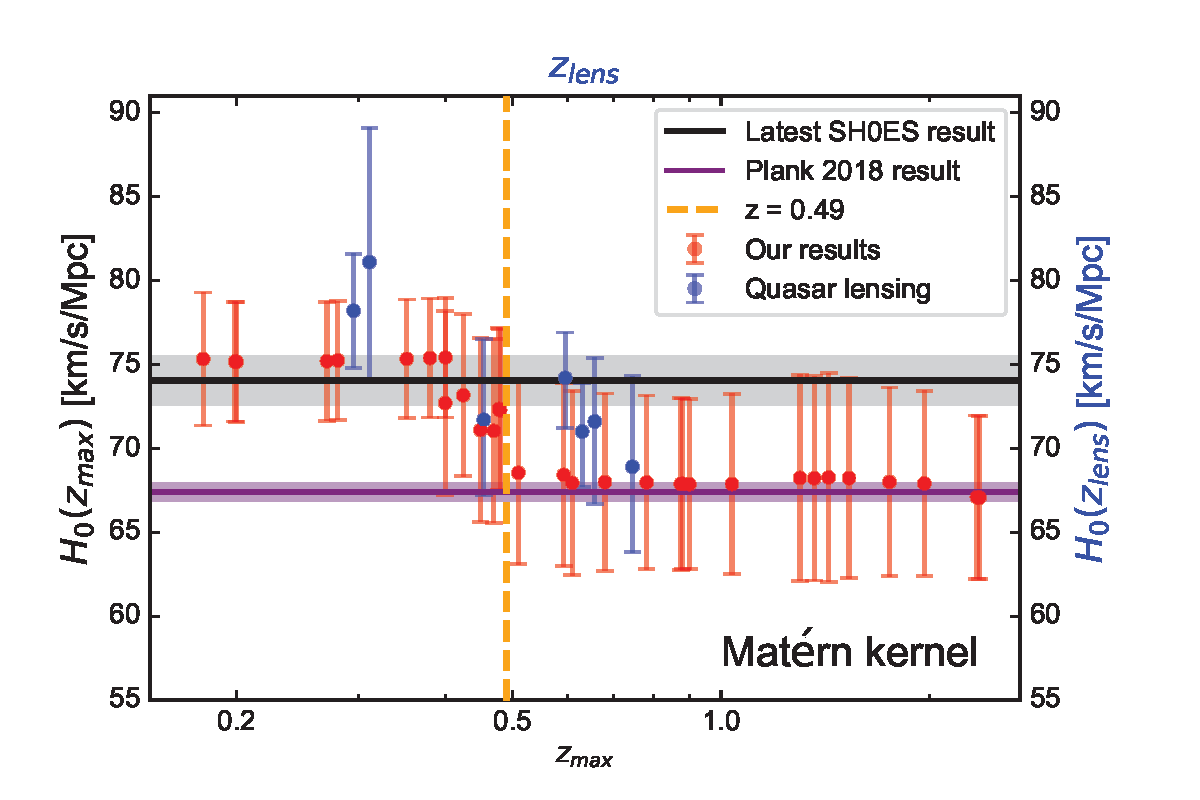}
\includegraphics[width=6.8 cm]{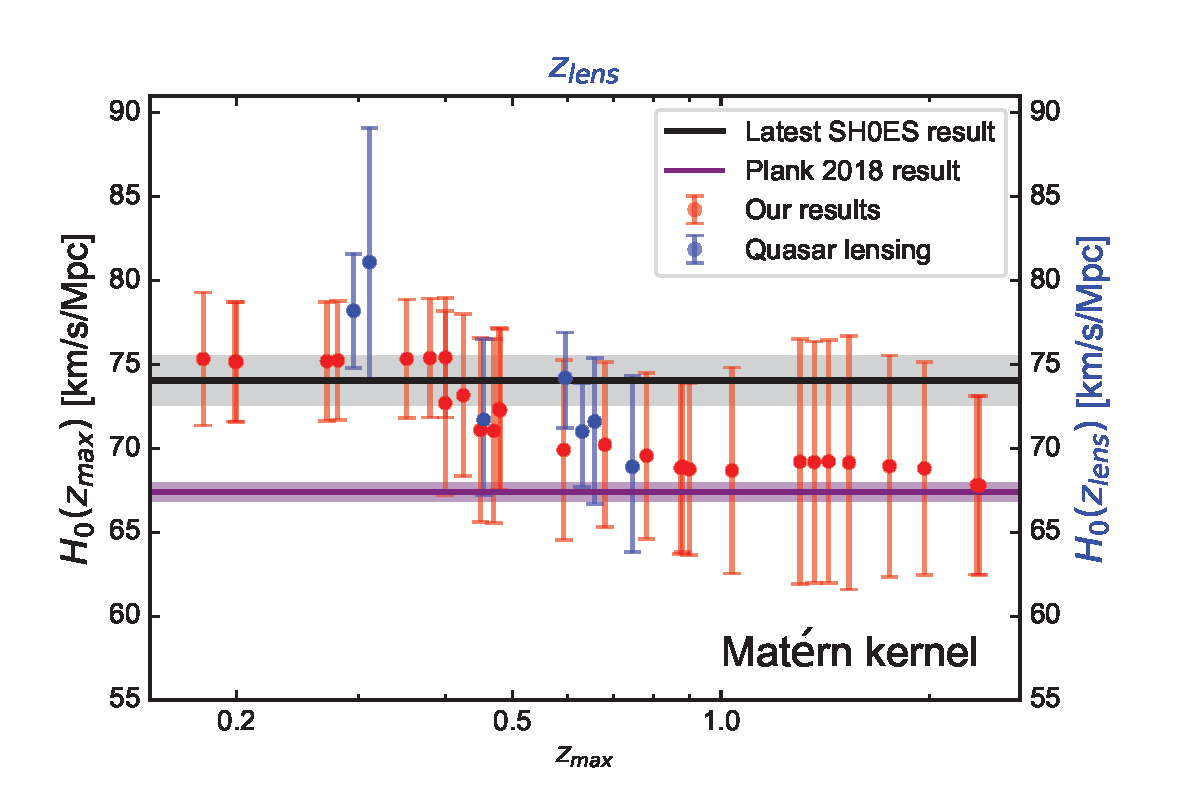}
\caption{Predictions of $H_{0}(z_{\rm max})$ derived from a sample of 36 $H(z)$ measurements (31 CC + 5 BAO). Here, $H_{0}(z_{\rm max})$ denotes the Hubble constant inferred from a dataset truncated at a maximum redshift $z_{\rm max}$. The red points represent the predicted values of $H_{0}(z_{\rm max})$. The gray and purple shaded regions indicate the constraints reported by the SH0ES and \emph{Planck} collaborations, respectively. The blue dotted vertical line marks the transition redshift at $z = 0.49$. Additionally, the blue points display the $H_0$ measurements derived from quasar lensing observations, plotted in the ($z_{\rm lens}$, $H_{0}(z_{\rm lens})$) plane. More details can be seen in Ref. \cite{2022MNRAS.517..576H}.
\label{fig:4.2-4}}
\end{figure}

Apart from investigating the evolution of $H_0$, there are some works attempting to provide an explanation for it. Combining the dark energy EoS with the Hubble constant, the results indicate that the variation of $w(z)$ appears to explain the evolution of the Hubble constant \citep{Jia2025_2}.
This idea simultaneously accounts for the deviation from the standard cosmological model observed in the current DESI BAO data, as well as the discrepancy between low-redshift and high-redshift data. Within the Friedmann-Lemaitre-Robertson-Walker metric, the Friedmann equations are 
\begin{equation}
    H^{2}=\left(\frac{\dot{a}}{a}\right)^{2}=\frac{8 \pi G}{3} \rho, \label{eq:Friedmann_1}
\end{equation}
\begin{equation}
    \frac{\ddot{a}}{a}=-\frac{4 \pi G}{3}(\rho+3 p), \label{eq:Friedmann_2}
\end{equation}
where $a=1/(1+z)$ is the scale factor, and $\dot{a}=da/dt$.
Combining equations (\ref{eq:Friedmann_1}) and (\ref{eq:Friedmann_2}), it can be found
\begin{equation}
    \rho=\rho_{0} \exp \left(-3 \int_{1}^{a}\left(1+w\left(a^{\prime}\right)\right) \frac{d a^{\prime}}{a^{\prime}}\right). \label{rho}
\end{equation}
The parameter $\rho_0$ denotes the cosmic total density at present. Substituting equation (\ref{rho}) into (\ref{eq:Friedmann_1}), we find 
\begin{equation}
    H_{0}^{2}=H(z)^{2} \times \frac{\rho_c}{\rho_0} (1+z)^{-3} \exp \left(-3 \int_{0}^{z} w\left(z^{\prime}\right) \frac{d z^{\prime}}{1+z^{\prime}}\right), \label{eq_H0z}
\end{equation}
where $\rho_c=3H_0^2/8\pi G$ is the critical density of the universe. This equation gives the relation between the equation of state $w(z)$ and the Hubble constant $H_0$. The above derivation only depends on the cosmological principle and Einstein's equations. Combining DESI BAO data with different SNe datasets, they derived the corresponding dark energy equation of state and the associated $H_0$, as shown in Fig. \ref{fig:4.2-7}. The results still exhibit a clear declining trend and are consistent with measurements at both low and high redshifts \citep{2022ApJ...934L...7R,2020A&A...641A...6P}. The method offers a possible explanation for the decline in $H_0$, suggesting that the variation in the equation of state leads to the decreasing Hubble constant. Additionally, \cite{2025MNRAS.536.3232M} attempted to explain the decline in $H_0$ using a void model. Their results show good agreement with the observed trend under both Gaussian and exponential void density profiles, suggesting that a local void structure could provide a viable physical mechanism to alleviate the Hubble tension while remaining consistent with a low background $H_0$ value.

In summary, the declining trend in $H_0$ was initially discovered through gravitational lensing observations. Subsequent studies, utilizing diverse datasets and methodologies, have consistently reproduced this declining trend. Although a mature physical model to explain this decline is not yet available, it offers a promising possibility for resolving the Hubble tension. Combined with constraints from FRBs, this holds promise for future studies on the evolution of $H_0$ \citep{2026ApJ...996...50K}.

\begin{figure}
    \centering
    \includegraphics[width=0.8\textwidth]{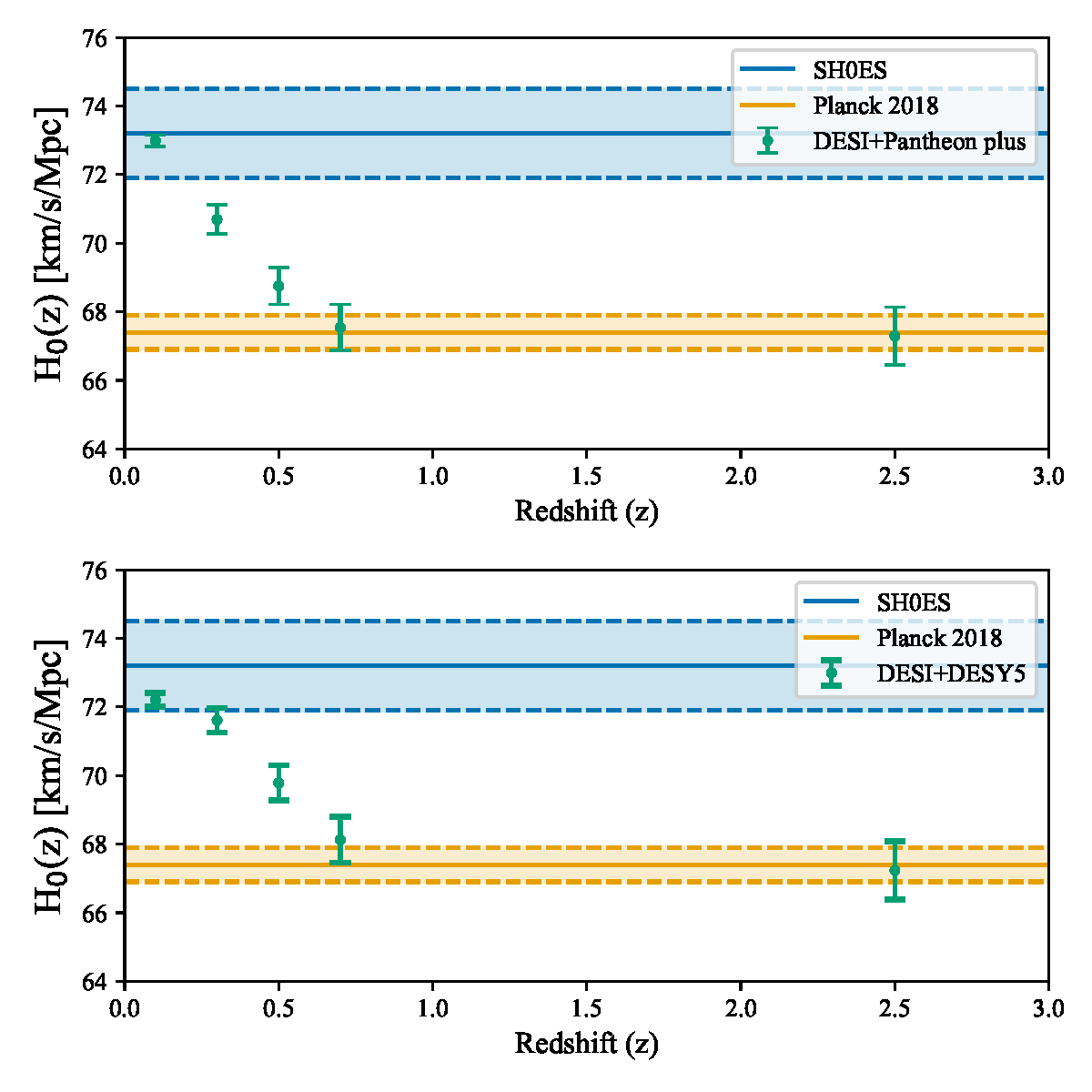} 
    \caption{The descending trend of the Hubble constant $H_0$ derived from the dark energy equation of state $w(z)$. Panel (a). Green data points represent the maximum a posteriori estimates of $H_0(z)$ with $1\sigma$ error bars, obtained from the combination of DESI DR2 BAO measurements and the Pantheon plus SNe Ia sample. The redshifts of these points correspond to the midpoints of their respective bins. The inferred $H_0$ is consistent with the local measurement within $1\sigma$ at low redshift and gradually converges toward the Planck CMB value at high redshift. This redshift dependence effectively mitigates the Hubble tension. Panel (b). Same as panel (a), but for the DESI DR2 BAO measurements and DESY5 SNe Ia sample. In both panels, local measurements of $H_0$ have not been used as a constraint. More details can be seen in Ref. \cite{Jia2025_2}.}
    \label{fig:4.2-7}
\end{figure}

\section{Future prospects}
As an emerging probe in cosmology, FRBs have demonstrated immense potential in recent years for addressing frontier scientific issues such as the Hubble tension and the EoS of dark energy. This paper reviews the developmental trajectory of FRB research, ranging from their initial discovery to the most recent investigation into the Hubble constant. Currently, the field of FRB research is at a critical transitional juncture, evolving from the early discovery phase toward a precision era driven by the accumulation of large samples.

Currently, existing radio telescope arrays, represented by CHIME, ASKAP, and FAST, have accumulated thousands of FRB samples and successfully localized hundreds of host galaxies. Leveraging this data, researchers are now capable of independently measuring the Hubble constant and placing preliminary constraints on dark energy using methods such as the Macquart relation, strong gravitational lensing. Notably, the exploration of the possible evolutionary trend of $H_0$ with redshift offers a fresh perspective on alleviating the current crisis faced by the $\Lambda$CDM model.

In the coming years, based on the observational capabilities of the CHIME instrument, the total number of observed FRBs is projected to reach $10,000$ \citep{2022A&ARv..30....2P,Zhang2023}. The SKA is expected to detect even more, with future estimates ranging from $10^4$ to $10^6$ events \citep{2017ApJ...846L..27F}. With the construction and commissioning of next-generation wide-field, high-sensitivity facilities, such as the SKA, DSA-2000 \citep{Hallinan2019DSA}, the Hydrogen Intensity and Real-time Analysis eXperiment (HIRAX) \citep{2022JATIS...8a1019C}, and CHORD \citep{2019clrp.2020...28V}, FRB observations will enter a brand-new era. We can expect to obtain a massive sample comprising tens of thousands of precisely localized FRBs. This will not only significantly reduce statistical errors, elevating FRBs into mature cosmological probes on par with SNe Ia, but will also empower us to delve into fundamental physical issues such as the baryon distribution in the IGM.





\clearpage

\begin{acknowledgements}
This work was supported by the
National Natural Science Foundation of China (grant Nos. 12494575, 12393812 and 12273009).
\end{acknowledgements}

\appendix                  

\bibliography{reference}
\bibliographystyle{raa}




\label{lastpage}

\end{document}